\documentstyle[prl,twocolumn,aps]{revtex}

\begin{document}

\twocolumn[\hsize\textwidth\columnwidth\hsize\csname
@twocolumnfalse\endcsname
\title{On the absence of spin-splitting in
$\alpha$-(BEDT-TTF)$_2$KHg(SCN)$_4$}
\author{N.~Harrison}
\address{National High Magnetic Field Laboratory, Los Alamos National 
Laboratory, MS-E536, Los Alamos, New Mexico 87545}
\author{N.~Biskup, J.~S.~Brooks and L.~Balicas}
\address{National High Magnetic Field Laboratory, Florida State 
University, Tallahassee, Florida 32310}
\author{M.~Tokumoto}
\address{Electrotechnical Laboratory, Tsukuba, Ibaraki 305, Japan}
\date{\today}
\maketitle

\begin{abstract}
We report the results of a detailed study of the 
field orientation-dependence of the de~Haas-van~Alphen 
waveform in 
$\alpha$-(BEDT-TTF)$_2$KHg(SCN)$_4$.
By considering the field 
orientation-dependence of the sign and phase of the fundamental 
$\alpha$ frequency, at fields 
both well above and below the kink transition field, it is 
found that a single value for the product of the effective mass with the 
electron {\it g}-factor can explain the experimental data deep within 
both the high magnetic field and low magnetic field phases.
This implies that spin-splitting does not 
occur within the low magnetic field phase until the angle between the 
magnetic field and the normal 
to the conducting planes is $\sim$~42$^\circ$.
This finding contrasts greatly with that recently published by Sasaki 
and Fukase, implying
that electron-electron interactions do not play a significant role 
in the formation of the charge-density wave ground state. 
The manner in which the amplitude of 
the waveform of the oscillations is damped within the 
low magnetic field phase is indicative of a non harmonically indexed 
reduction of the amplitude, thereby eliminating both 
magnetic breakdown and impurity scattering as dominant 
damping mechanisms 
within this phase. Meanwhile, the presence of a large 
amplitude second harmonic within the low magnetic field phase that 
has a negative sign over a broad range of angles can only be 
explained by the frequency doubling effect.
\end{abstract}

\pacs{XXXX}
]\narrowtext

\section{introduction}
Charge-transfer salts of the composition 
$\alpha$-(BEDT-TTF)$_2M$Hg(SCN)$_4$ (where $M=$~K, 
Tl or Rb) exhibit a complex phase diagram that is profusely sensitive both
to temperature $T$ and the orientation of an applied magnetic 
field ${\bf B}$ 
\cite{tokumoto1,sasaki1,kartsovnik1,qualls1,harrison1}.
The observation of numerous magneto-oscillatory 
phenomena, as a function of the magnitude and orientation of ${\bf B}$ 
\cite{kartsovnik2,harrison2}, has firmly established the existence of a 
reconstructed Fermi surface below the transition temperature $T_{\rm p}$, 
occurring at $\sim$~8~K 
in the $M=$~K and Tl salts \cite{tokumoto1} 
(or $\sim$~10~K in the $M=$~Rb salt). 
Direct evidence for a lattice superstructure, which would be 
necessary in order to unambiguously distinguish a charge-density wave 
(CDW) ground state from one that is a spin-density wave (SDW)
\cite{gruner1}, however, has not been forthcoming. Only 
very recently has sufficient indirect evidence been accumulated so as 
to tip the balance of the arguments in favour of explanations 
involving CDW ground states. 
Notably, antiferromagnetism is either weak \cite{sasaki2,pratt1} 
or absent \cite{miyagawa1}, and $T_{\rm p}$
is strongly suppressed by the application of a magnetic 
field \cite{kartsovnik1,qualls1,harrison1,kovalev1}.
The physics of this material then changes abruptly on passing 
through the first order ``kink'' transition field \cite{osada1}, 
$B_{\rm k}$ (occurring at $\sim$~23~T in the $M=$~K salt),
above which it becomes diamagnetic \cite{harrison1}. 
While the experimentally delineated phase boundaries are consistent 
with this being a transition from a commensurate 
CDW$_0$ phase into a high magnetic field modulated CDW$_x$ phase 
\cite{kartsovnik1,qualls1,harrison1}, as predicted
by several recent theoretical models \cite{zanchi1,mckenzie1}, 
the physical changes incurred experimentally are difficult to reconcile 
with such a simple transition \cite{harrison1}. 

Quantum oscillations have proven
to be especially sensitive to $B_{\rm k}$
\cite{osada1}. All signatures of a reconstructed Fermi surface are 
lost at high magnetic fields \cite{sasaki1,harrison1,caulfield1,house1}, 
while the effective mass $m^\ast$ of the dominant $\alpha$ frequency
appears to increase \cite{sasaki1,harrison3,christ1,uji1}. 
Most notable, perhaps, is the change in the physical appearance of the 
waveform from one that is strongly damped but displaying split 
maxima at fields below $B_{\rm k}$ \cite{sasaki2,tokumoto2,sasaki3}, 
to one that is triangular at fields above $B_{\rm k}$ 
\cite{harrison1,harrison4,honold1}. 
The origin of the splitting of the waveform within the 
CDW$_0$ phase is, itself, a contemporary issue. Numerous 
publications \cite{sasaki2,tokumoto2,sasaki3} attest to the fact that this 
splitting
appears to resemble the ``spin-splitting'' phenomenon that occurs when 
the degree to which the Landau levels are split by the Zeeman energy, 
$\Delta\varepsilon=g\hbar eB/m_{\rm e}$ ($g$ being the electron {\it 
g}-factor and $m_{\rm e}$ being the free electron mass), becomes 
commensurate with an odd half-integer multiple of the cyclotron 
energy, $\hbar\omega_{\rm c}=\hbar eB/m^\ast$ \cite{shoenberg1}. 
The observation of a split waveform does not, however,
prove the existence of spin-splitting. A similar split 
waveform occurs in the CDW compound NbSe$_3$ \cite{monceau1}, 
for example, yet this 
was recently shown not at all to be related to the Zeeman effect 
\cite{harrison5}. The frequency doubling (FD) effect 
\cite{harrison6}, which 
occurs when an additional term proportional to the square of the 
oscillatory 
chemical potential $\tilde{\mu}$ modulates the free energy of a CDW 
ground state, could provide an alternative explanation. In spite of 
the 
fact that seemingly strong arguments have been given for 
spin-splitting in the $\alpha$-(BEDT-TTF)$_2M$Hg(SCN)$_4$ salts 
\cite{sasaki2,tokumoto2,sasaki3,sasaki4}, this hypothesis has not 
actually
been thoroughly tested.

The purpose of the present paper is to thoroughly investigate the 
possibility of both the spin-splitting and FD effects in 
$\alpha$-(BEDT-TTF)$_2$KHg(SCN)$_4$ by performing a 
detailed investigation of the field orientation-dependence of the 
waveform at a variety of magnetic fields 
both above and below $B_{\rm k}$. In this work, the product 
$\nu^\ast_0g$ of the 
degree to which the effective mass is enhanced 
$\nu^\ast_0=m^\ast_0/m_{\rm e}$ with the electron {\it g}-factor is 
determined by fitting to the field orientation of the fundamental oscillation
sign and phase. As will be discussed in Section III, this is required 
in order to 
determine the relative importance of electron-electron ({\it e-e}) and 
electron-phonon ({\it e-ph}) interations in the formation of the 
ground state. The validity of the canonical ensemble for describing 
the field-orientation dependence of the waveform within the high 
magnetic field phase is discussed in Section IV, while the details 
concerning the anomalous 
behaviour of the quantum oscillations within the low magnetic field 
phase are described in Section V. We turn to a discussion of the 
frequency doubling effect 
in Section VI and summarise the results in Section VII.
\section{experimental}
The single crystal sample of $\alpha$-(BEDT-TTF)$_2$KHg(SCN)$_4$
of volume $\sim$~0.8~mm$^3$, used in this study,
was the same as that used for the magnetic torque measurements in 
Reference \cite{harrison1}.
It was mounted on the moving plate of a phosphor bronze
capacitance cantilever, which was itself attached to a rotating 
platform for which the axes of torque and rotation were 
parallel to each other, yet both perpendicular to ${\bf B}$.
The angle between ${\bf B}$ and the normal to the capacitance 
plates was approximately the same as the angle $\theta$ between 
${\bf B}$ and the normal ${\bf n}$ to the conducting planes of the 
sample. The capacitance, of order $\sim$~1~pF,
was measured by means of a capacitance bridge energized with 
30~V~rms at 5~KHz and
was observed to change by less than 0.1~\%. 
Since this implies a maximum angular 
displacement of $\sim$~$\frac{1}{20}^\circ$, torque interaction 
effects were insignificant.
Static magnetic fields extending to $\sim$~32~T were provided 
by the National High Magnetic Field Laboratory, Tallahassee, while 
a constant temperature of $\sim$~450~mK was obtained using 
a $^3$He refrigerator. 

Because the interlayer transfer integral $t_\bot$ of 
$\alpha$-(BEDT-TTF)$_2M$Hg(SCN)$_4$ charge-transfer salts is 
immeasurably small compared to those within the planes, these 
materials provide some of the best known examples of ideal 
multilayered
two-dimensional (2D) Fermi liquids \cite{harrison4}. The only 
significant 
component of the Landau diamagnetic susceptibility is 
that projected along ${\bf n}$. Because
$\mathbf{\tau}=\mathbf{M}\times\mathbf{B}$,
the oscillatory componenent of the oscillatory magnetic torque is
\begin{equation}\label{torque}
    \tilde{\tau}_\theta=-\tilde{M}_{\bot,\theta} B\sin{\theta},
\end{equation}
where $\tilde{M}_{\bot,\theta}$ is the oscillatory component of 
$\mathbf{M}$ parallel to ${\bf n}$.
\section{Field orientation-dependence of the dHvA phase}
Examples of the oscillatory magnetic torque of
$\alpha$-(BEDT-TTF)$_2$KHg(SCN)$_4$, measured 
in static magnetic 
fields of up to 32~T and at several different field orientations, 
are shown in Fig. \ref{wiggles}. Note that these data
closely resemble earlier measurements made on the same material
\cite{christ1,uji1}. A Fourier transformation of the data over a 
restricted 
range of $B$ (in the $1/B$ domain) within the low magnetic field 
phase at 
$\theta\sim$~8.8$^\circ$, shown in Fig. \ref{transforms}, reveals 
a plethora of harmonics indicative of a good sample quality.

A number of recent articles have shown that the oscillations of 
$\tilde{\mu}$ (in this and other 2D materials) invalidates a simplistic 
analysis of the dHvA oscillation data in terms of the  
Lifshitz-Kosevich (LK) formula \cite{harrison4,harrison7}. The 
reasons 
for this are twofold. First, the LK formula is suited only to systems 
in 
which the Fermi surface is significantly curved in all three 
$k$-spatial 
dimensions \cite{harrison4,shoenberg1}. Second, the oscillations of 
the chemical potential $\tilde{\mu}$ 
significantly perturb the waveform of the oscillations so as to cause the 
amplitude and sign of each of the $p>$~1 harmonics to depart 
significantly 
from those predicted by the LK model \cite{harrison4}. 
Further complications may also arise as a result of 
interactions involving the condensate itself, due either to magnetic 
breakdown \cite{sasaki5}, the FD effect \cite{harrison6} or induced 
currents. Induced currents, that contribute an additional oscillatory 
structure 
to the dHvA waveform, have now been shown to occur both in static 
magnetic fields \cite{harrison1} and pulsed magnetic fields 
\cite{honold1,harrison8}.

In spite of the fact that the waveform of the dHvA oscillations is 
significantly perturbed by $\tilde{\mu}$ in the canonical ensemble, 
the underlying sign 
and phase of the fundamental frequency (which we shall label as $p=$~1) 
is the 
same as that in the 
grand canonical ensemble (or equivalently the LK model) for 
which $\tilde{\mu}$ is assumed to be 0 
\cite{harrison4,shoenberg1}. The amplitude of the fundamental 
oscillations in the magnetic torque can therefore be written in the 
form
\begin{equation}\label{phaseonly}
    \tilde{\tau}_{1,\theta}\approx
    A_{1,B,T,\theta}
    \sin{\bigg(\frac{2\pi F}{B}\bigg)}
    S_{1,\theta}\sin{\theta},
\end{equation}
where $A_{1,B,T,\theta}$ is a monotonically varying sign and phase 
independent prefactor (for which there is no simple algebraic form in 
the canonical ensemble) and $F$ is the dHvA frequency.
Only the Zeeman term $S_{1,\theta}=\cos(\pi\nu^\ast_\theta g/2)$
determines the sign and phase of the oscillations \cite{shoenberg1}, 
for which 
$\nu_\theta^\ast=m^\ast_\theta/m_{\rm e}=m^\ast_0/m_{\rm e}\cos{\theta}$ 
\cite{wosnitza1}.

The magnitude $|S_{1,\theta}|$ becomes unity 
whenever $\Delta\varepsilon$ becomes commensurate with 
$\hbar\omega_{\rm c}$, or, equivalently, when the product 
$\nu_\theta^\ast g$ becomes equal to an even integer.
Conversely, 
whenever $\nu_\theta^\ast g$ is equal to an odd integer, the 
amplitude of the fundamental oscillation frequency undergoes a 
node (often called a `spin-splitting zero') \cite{shoenberg1}. 
The $\theta$-dependence of $\nu_\theta^\ast$ then causes 
the amplitude of the 
fundamental to pass through a series of spin-splitting zeros
upon rotation of the sample in a magnetic field.
The experimentally determined 
positions of these nodes can then be identified with 
particular values of $\nu^\ast_0 g/\cos{\theta}$, enabling an 
accurate estimate of $\nu^\ast_0 g$ to be made.

A study of this type was recently made by Sasaki and Fukase at 
various magnetic fields both above and below $B_{\rm k}$ in 
$\alpha$-(BEDT-TTF)$_2$KHg(SCN)$_4$ \cite{sasaki4}. 
The interpretation of the positions of the nodes is not, however, entirely
unambiguous \cite{shoenberg1}. Since the first spin-zero could 
correspond to any odd integer value of 
$\nu^\ast_0 g/\cos{\theta}=$~1, 3, 5, 7\ldots, 
part of the investigation involves determining which of these it is 
likely to be. The process of distinguishing these becomes trivial only
when many nodes are observed. For this reason, our results within the 
CDW$_x$ phase at $B\sim$~26.5~T
({\it i.e.} for the interval in $B$ between 23.0 and 31.253~T), 
presented in Fig. \ref{phase}a, are in excellent agreement with those 
of 
Sasaki and Fukase \cite{sasaki4}.
For clarity, this data is reproduced in Fig. \ref{phase}b, together 
with a 
solid line representing the functional form of 
$S_{1,\theta}\sin\theta$ best 
able to reproduce the correct phase of the oscillations and 
positions of 
the spin-zeros. Note that the solid line is a fit only to the sign 
of the 
oscillations and not to the amplitude, yielding
$\nu^\ast_0 g=$~3.67~$\pm$~0.02.

Also in agreement with Sasaki and Fukase \cite{sasaki4}, 
in Fig. \ref{phase}a we see that the apparent angular positions of 
the nodes appear 
to shift on lowering $B$ through the transition field $B_{\rm k}$. 
However, in disagreement with the results of Sasaki and Fukase, we 
observe that, at much lower fields, the positions of the nodes 
eventually shift 
back to the same positions as those at $\sim$~26.5~T. 
Therefore, in 
contrast to Sasaki and Fukase \cite{sasaki4}, we 
find that a single value of 
$\nu^\ast_0 g\sim$~3.67 is able to fit the field 
orientation-dependence of 
the sign of the fundamental oscillation amplitude deep within both 
the 
CDW$_0$ and CDW$_x$ phases. 
To illustrate this point more clearly, in Fig. \ref{phase}c we have 
replotted the field orientation-dependence of the 
oscillation amplitude in the magnetic torque at 16.5~T 
({\it i.e.} for $B$ between 15.0 and 18.2~T),
together with a solid line representing the functional form of 
$S_{1,\theta}\sin\theta$, with $\nu^\ast_0 g\sim$~3.67 as within the 
CDW$_x$ phase. Clearly, this value of $\nu^\ast_0 g$ is able to 
reproduce 
the positions of the nodes quite adequately. In contrast, when
the value of $\nu^\ast_0 g\sim$~4.7 quoted by Sasaki and 
Fukase \cite{sasaki4} is inserted into $S_{1,\theta}\sin\theta$, as 
indicated by the dashed line, the positions of the nodes are not 
accurately reproduced. 

There are at least
two notable flaws with the analysis of Sasaki and Fukase 
\cite{sasaki4}. First, they assume 
the index of the first node to switch abruptly from 
$\nu^\ast_0 g/\cos{\theta}=$~5 to $\nu^\ast_0 g/\cos{\theta}=$~7
at $B_{\rm k}$. Second, their reported value of $\nu^\ast_0 g=$~4.7 
within the CDW$_0$ phase requires the existence of a 
node at $\theta\sim$~20$^\circ$ that has never actually been 
observed. 
The data in Fig. \ref{phase} shows no evidence for either an abrupt 
change 
in $\nu^\ast_0 g$ or the development of a node at 
$\theta\sim$~20$^\circ$ 
on passing into the CDW$_0$ phase. Sasaki and Fukase did not attempt to 
verify the presence or absence of a node at this orientation
\cite{sasaki4}.

A reliable extraction of $\nu^\ast_0 g$ is required for determining the 
relative importance of {\it e-e} and  {\it e-ph} interactions. 
According to Fermi liquid theory, 
{\it e-e} and {\it e-ph} interactions affect $\nu^\ast_0$ and $g$ 
differently 
\cite{shoenberg1}. In the case of {\it e-e} interactions, an increase 
in 
$\nu^\ast_0$ is approximately offset by a reduction in $g$ so that 
there 
is no overall change in the product $\nu^\ast_0 g$. This is not, 
however, 
the case with {\it e-ph} interactions. We can therefore conclude 
that, 
since $\nu^\ast_0 g$ does not change appreciably in this experiment, 
there is no significant change in {\it e-e} interactions on passing 
between the 
CDW$_0$ and CDW$_x$ phases, thereby contrasting with
the conclusion reached by Sasaki and Fukase \cite{sasaki4}. Either the 
effective electron density is not a significant factor in determining 
the relative strengths of the effective Coulomb interaction between 
the two regimes, or, alternatively, {\it e-e} interactions do not 
play a 
significant role in the formation of the CDW ground state. This should be 
of no surprise \cite{gruner1}, given that 
CDW ground states are commonly thought to involve {\it e-ph}
interactions rather than {\it e-e} interactions. 

One important question that remains to be answered, therefore, is 
whether the apparent change in the effective mass of the $\alpha$ 
frequency on crossing $B_{\rm k}$ is genuinely related to a change in 
the strengh of the {\it e-ph} interactions \cite{sasaki1,harrison3,christ1,uji1}, or 
whether it is an artefact of the temperature dependence of the 
condensate \cite{sasaki5}. What complicates matters futher is that 
the effective mass estimates have a history of being 
unreliable in this material \cite{sasaki1,harrison3,christ1,uji1,honold1}. 
Within the CDW$_0$ phase, for example, different values of $m^\ast$ 
are obtained depending on whether one analyses 
Shubnikov-de~Haas (SdH) or dHvA data \cite{harrison3}. Under normal 
circumstances, 
dHvA data is more reliable owing to the fact it is 
derived from a thermodynamic function of state. However,
there also exists the possibility that gaps of order $2\Delta$ in 
the energy spectrum resulting from the formation of the CDW state,  
lead to breaks in the $\alpha$ orbit trajectory that 
then have to be overcome by magnetic breakdown in a magnetic field 
\cite{uji1}. It has been argued that since $2\Delta$ falls with 
increasing temperature, this should lead to an 
additional temperature-dependent term in the quantum oscillation 
amplitude that could 
potentially cause the effective mass within the CDW$_0$ phase to appear 
artificially low \cite{sasaki5}. In Section V, however, 
we will show that magnetic 
breakdown appears not to be the dominant form of damping within the 
CDW$_0$ phase.
In any case, the 
reported difference in the degree to which the effective mass is 
enhanced between the CDW$_0$ and CDW$_x$ phases, 
$\delta\nu^\ast_0\sim$~0.5, appears to be quite significant. 
Within the CDW$_0$ phase, dHvA measurements all agree that 
$\nu^\ast_0\sim$~1.5~$m_{\rm e}$ 
\cite{sasaki1,harrison3,christ1,uji1,house2}. 
Within the CDW$_x$ phase, however, only one estimation of
$\nu^\ast_0\sim$~2.0~$m_{\rm e}$ has been made that properly 
accounts for the effects of induced currents, now shown to occur 
both in static and pulsed magnetic fields \cite{harrison1}. Since
$\alpha$-(BEDT-TTF)$_2$KHg(SCN)$_4$ is believed to possess 
a CDW ground state, of some form 
\cite{kartsovnik1,qualls1,harrison1,miyagawa1,mckenzie1}, changes in 
$\nu^\ast_0$ between CDW sub-phases could be expected. 
A common observation in all CDW materials is 
that gaps open in the phononic density of states as well as in the 
electronic density of state, often referred to as the Kohn anomaly 
\cite{gruner1}. Since the net {\it e-ph} coupling strength is 
determined by an integration over both the phononic and electronic 
densities of states \cite{shoenberg1}, an increase in the 
effective mass should be expected on passing into a phase within 
which $2\Delta$ is lower. 

This still leaves the behaviour within the narrow field region, 
18.2~$<B<$~20.3~T, unexplained.
One likely possibility is that the analysis procedure is not able to 
separate contributions to the dHvA effect originating from the 
different CDW$_0$ and CDW$_x$ phases over this range, owing to the 
possible existence of one or more first order changes in the electronic 
structure just below $B_{\rm k}$ \cite{harrison6}. We also note that 
at 
higher angles, $|\theta|\gtrsim$~45$^\circ$, another phase, CDW$_y$, 
has been proposed to exist \cite{qualls1}, which could only 
complicate matters further.
\section{dHvA waveform within the CDW$_x$ phase}
That the conventional form of $S_{1,\theta}$ is well obeyed 
both in the CDW$_0$ and CDW$_x$ regimes of
$\alpha$-(BED-TTF)$_2$KHg(SCN)$_4$,  
(with the exception of a narrow field interval 
immediately below $B_{\rm k}$), implies that this material has, at all 
times, a well defined set of Landau levels characteristic of a normal 
Fermi liquid in a magnetic field. Given, also, that the product $\nu^\ast_0g$ 
is close to an integral value ({\it i.e.} 4), it is of no surprize that the high 
magnetic field phase closely resembles a canonical ensemble of 
electrons for which the spins are approximately degenerate 
\cite{harrison4}. On inserting more exact parameters, 
$\nu^\ast\sim$~2.0~ \cite{honold1}, 
$\gamma\sim$~0.68 \cite{harrison1,note1} and $\nu^\ast_0g\sim$~3.67 
(this work) into the numerical model of Reference \cite{harrison4}, the 
canonical ensemble (calculated in Fig. \ref{waveform}b) is able to 
reproduce the 
experimentally observed magnetic torque in Fig. \ref{waveform}a rather 
well. The data in Fig. \ref{waveform}a was taken in a dilution fridge at 
$T\sim$~27~mK and $\theta\sim$~7$^\circ$. Note that the parameters 
$F$, $\nu^\ast_0$, $\gamma$ and $\nu^\ast_0g$ are constants specific to 
the material that have been determined experimentally and cannot 
be arbitrarily adjusted as fitting parameters. Only the scattering rate
$\tau^{-1}\sim$~0.6$\times$10$^{12}$~s$^{-1}$, which is always
sample-dependent, can be adjusted in order to obtain the best representation 
of the experimental data.

These same parameters, when inserted into the numerical canonical 
ensemble calculation, are also able to reproduce the correct field 
orientation-dependence of the fundamental ($p=1$) amplitude of the 
magnetic torque in Fig. \ref{ampsCDWx}a, at least for 
$|\theta|\lesssim$~45$^\circ$. The same numerical model also predicts 
the correct sign of the second ($p=2$) harmonic, albeit that the field 
orientation-depedence of its amplitude is less accurately reproduced. 
By ``sign'' we refer to the sign of $a_p$ that correctly depicts the 
waveform of the oscillations when it is decomposed into a Fourier 
expansion of the form
\begin{equation}\label{fourierexpansion}
    \tilde{M}\approx\sum\limits_p
    \frac{a_pN\beta^\ast}{\pi p}\sin\bigg(\frac{2\pi pF}{B}\bigg).
\end{equation}
Here,
$\beta^\ast=\hbar e/m^\ast$ is the double Bohr magneton, $N$ is the 
density of carriers giving rise to the $\alpha$ frequency quantum 
oscillations and $|a_p|<1$ represents the degree to which the amplitude of 
each harmonic is attenuated due to the combined effect of impurities, 
spin and temperature. In the canonical ensemble, 
there is no simple way to separate each of these contributions 
\cite{harrison4}. 
The field orientation-dependence of the sign of the second 
harmonic is expected to remain positive in the canonical ensemble 
for all angles when 
$\gamma>$~0.5 \cite{harrison4}
({\it i.e.} when the density of states of the 2D Fermi surface 
pocket is larger than that of the quasi-one-dimensional sheets). 
In Fig. \ref{ampsCDWx}b we can see that,
while the grand canonical ensemble ({\it i.e.} the LK model which 
assumes a fixed chemical potential) is equally well able to explain the 
behaviour of the fundamental, it fails to account for the positive 
sign of the second 
harmonic. This illustrates the hazards associated with 
fitting the LK model to a 2D system for which it does not apply 
\cite{harrison3,christ1,uji1,fitting}. In Section V we will show 
that this issue becomes particularly important when attempting to 
understand the oscillations within the CDW$_0$ 
phase.

In spite of the fact that the models are able to predict the 
correct form of the fundamental amplitude of the dHvA oscillations 
at small angles, they cannot 
account for their rapid attenuation at larger 
angles, $|\theta|\gtrsim$~45$^\circ$, in Fig. \ref{ampsCDWx}. One 
possible explanation   
is that the scattering rate $\tau^{-1}({\bf k})$ is strongly 
dependent on ${\bf k}$, with ``hot spots,'' or possibly even ``hot 
bands,'' occurring at certain values of $k_z$ (the lattice vector 
parallel to ${\bf n}$) \cite{harrison9}. Such effects 
have been suggested to be important in some of the Bechgaard salts
\cite{chaikin1}. Since a dHvA experiment senses only a weighted 
average of $\tau^{-1}({\bf k})$, the number of orbits that 
intersect hot regions of the Fermi surface could increase for large $\theta$. 
It was noted in Reference \cite{symington1}, that the experimentally 
observed scattering rate appears to increase roughly in proportion 
to $\tan\theta$. This damping of the oscillations at large angles 
will be the subject of a future publication \cite{harrison9}.
\section {dHvA waveform within the CDW$_0$ phase}
Thus far, we have shown that at least two parameters associated with the 
$\alpha$ frequency appear not to change on 
traversing $B_{\rm k}$; namely, 
its fundamental frequency $F\sim$~670~T and the product $\nu^\ast_0g$. 
The same cannot be said with confidence about the degree to which 
the effective mass is enhanced $\nu^\ast_0$, the $\gamma$ parameter
(which quantifies the fraction of the density of states occupied by the
2D pocket), or 
the scattering rate $\tau^{-1}$. A number of groups have reported an 
apparent increase in the scattering rate within the low magnetic field 
CDW$_0$ phase with respect to that within the high magnetic field phase
\cite{sasaki1,harrison3,christ1}. Others have attributed the loss of 
amplitude of the $\alpha$ frequency within the CDW$_0$ phase to 
magnetic breakdown effects \cite{uji1,sasaki5}. While the latter might 
be expected following the introduction of an additional periodic 
potential $2\Delta_0$ within the CDW$_0$ phase
\cite{kartsovnik2,harrison2}, neither of these two possibilities 
can satisfactorily explain the experimental 
data. For either of them to be true, the field dependence of the amplitude 
of each harmonic $p$ (having corrected for its temperature 
dependence) would have to be proportional to $R_{p,B}\approx
\exp(-\pi p/\omega_{\rm c}\tau-pB_0/B)$. The first term within the 
exponent accounts for scattering due to impurities \cite{shoenberg1} while 
the second accounts for magnetic breakdown, having assumed that no 
Bragg reflection takes place on the $\alpha$ orbit; as is commonly 
assumed \cite{kartsovnik2,harrison2}. Because the field
dependence of both of these terms is the same, there is no way to 
distinguish them experimentally. We can therefore write this in the more 
generic form,
$R_{p,B}\approx\exp(-p\Upsilon/B)$, where $\Upsilon$ represents the 
total degree of damping inclusive of both effects. In Fig. \ref{ampsCDW0}a,
the experimentally observed fundamental amplitude of the oscillations 
within the low magnetic field phase, at $B\sim$~16.5~T,
can be approximately reproduced using the numerical model 
by setting $\Upsilon\sim$~79~T. This is
equivalent to a scattering rate of 
$\tau^{-1}\sim$~2.9$\times$10$^{12}$~s$^{-1}$, 
comparable to that 
obtained in Reference \cite{harrison3}, or, alternatively, to a
characteristic magnetic breakdown field of $B_0\sim$~79~T. 
Implicit to either of these explanations, however, is the reduction 
of the amplitude of the second ($p=$~2) harmonic with respect to that of the 
fundamental by another factor of approximately 
$R_{B}\approx\exp(-\Upsilon/B)\sim$~10$^{-2}$. This appears not 
the case experimentally, however.
For example, when we 
calculate the waveform using the numerical model
(with $\Upsilon\sim$~79~T), the amplitude of the 
second harmonic in Fig. \ref{ampsCDW0}a is 
roughly two orders of magnitude smaller than that detected 
experimentally. This would also be the case were we to calculate the 
waveform using the grand canonical ensemble, or were we to take
into consideration FD effects (see Section VI).
Clearly, the presence of a second harmonic with an amplitude 
that is measured to be an appreciable fraction of that of the fundamental is 
inconsistent with a harmonically indexed damping factor of 
the form $R_{p,B}\approx\exp(-p\Upsilon/B)$ with $\Upsilon$ being as 
large as 79~T.
We can therefore eliminate both impurity scattering and magnetic breakdown as 
dominant mechanisms for the damping of the the dHvA oscillations 
observed within the CDW$_0$ phase, since both of these lead to
harmonically indexed damping factors. The only alternative 
explanation, therefore, is that the quantum oscillations within the 
CDW$_0$ phase is uniformally suppressed in amplitude in 
a manner that not indexed 
to the harmonics. 
Evidence for a non harmonically indexed reduction of 
the amplitude has already been published \cite{harrison3}.
In Reference \cite{harrison3}, the Dingle plots of the fundamental and 
second harmonic were found to have approximately the same slope, 
indicating the existence of a amplitude reduction factor that is not indexed to $p$.
In order to account for these experimental observations, we 
can notionally introduce a damping factor of the form
$R^\prime_{B}\approx\exp(-\Upsilon^\prime/B)$ within the CDW$_0$ 
phase that is independent of 
$p$ but that operates in addition to the conventional damping that occurs within the high 
magnetic field phase. An exponential form for $R^\prime_{B}$ is 
required in order to account for the fact that the Dingle plots 
are linear \cite{harrison3}.

The most trivial interpretation of a non harmonically indexed damping 
factor is that where the effective volume of the sample 
contributing to the dHvA signal is reduced by a factor
$R^\prime_{B}\approx\exp(-\Upsilon^\prime/B)$. A volume reduction 
factor of this type could, for example, be expected  
were the CDW$_0$ phase composed of two coexisting phases 
spatially separated over 
distances larger than the cyclotron length, only one phase of which 
yields dHvA oscillations of the $\alpha$ frequency, with their
composition then changing with field. 
When the dHvA 
waveform in Fig. \ref{ampsCDW0}b is calculated using the same material
parameters as within the CDW$_x$ phase, but with an additional 
empirical damping term of the form 
$R^\prime_{B}\approx\exp(-\Upsilon^\prime/B)$, 
setting
$\Upsilon^\prime\sim$~60~T, the model is able to reproduce the 
experimentally observed amplitudes
somewhat better than in Fig. \ref{ampsCDW0}a. In particular, the 
model now predicts the second harmonic to have the correct order of 
magnitude, albeit that is has the opposite sign to that observed 
experimentally. We will return to a discussion of 
the sign of the second harmomic in Section VI where we consider FD 
effects.

Above we have shown that neither impurity scattering nor magnetic breakdown 
can account for the strong damping within the CDW$_0$ phase. To carry 
this argument further, 
it can be shown that both of these explanations 
are unphysical for other reasons. For example, a 
scattering rate is usually determined by the number of defects and 
impurities in a metal, and this number is not expected to change across a 
phase transition. The product $m^\ast_0\tau^{-1}$ invariably remains 
constant. Similarly, the estimated value of $\Upsilon\sim$~79~T 
significantly exceeds the magnetic breakdown field 
$B_0\approx n\varepsilon_{\rm gap}^2B/2\varepsilon_{\rm 
F}\hbar\omega_{\rm c}\sim$~20~T \cite{shoenberg1} that should be 
expected for $n\sim$~6 
magnetic breakdown nodes of size $\varepsilon_{\rm gap}\approx 
2\Delta_0\approx$~4~meV \cite{harrison1}; 
$\varepsilon_{\rm F}=\hbar eF/m^\ast$ being the Fermi energy.
\section{frequency doubling}
The most distinguishing feature of the oscillations in the magnetic 
torque within CDW$_0$ phase is the presence of a strong second harmonic. 
The ratio of the harmonics is 
unaffected by the uniform non harmonically indexed reduction in the
amplitude of the oscillations discussed in the preceding section. 
Another important feature 
of the dHvA oscillations within the CDW$_0$ phase, which has not 
been addressed by earlier publications, is that the sign of the 
second harmonic is negative compared to one that is positive above 
$B_{\rm k}$. The change in sign of the second harmonic between the 
low and high magnetic field phases gives rise to a 
node at $B_{\rm k}$ as observed by Uji {\it et al.} \cite{uji1}.

The negative sign of the second harmonic within the CDW$_0$ phase is 
clearly unexpected in the canonical ensemble. It is also 
inconsistent with spin-splitting in the grand canonical ensemble (or 
LK model), for which a positive sign should also be expected. In fact, 
the negative sign of the second harmonic over a wide range of angles, 
0$^\circ<|\theta|<$~42$^\circ$, is inconsistent with any model of the 
dHvA effect. As shall become clear below, it is, however, expected to be 
negative when frequency doubling effects are taken into consideration.

In order to model the extent to which the FD effect can affect 
the waveform, it is useful to consider the proportionality 
$\tilde{\mu}=B\tilde{M}/N$ \cite{harrison4,itskovsky1} which, when 
combined with Equation (\ref{fourierexpansion}), enables
the oscillations in the chemical potential to be written as a series 
expansion of the form
\begin{equation}\label{chemp}
    \tilde{\mu}\approx\sum\limits_p 
    \frac{a_p\hbar\omega_{\rm c}}{\pi p}\sin\bigg(\frac{2\pi pF}{B}\bigg).
\end{equation}
According to the frequency doubling model, oscillations in the 
chemical potential give rise to an additional term in the free energy 
of the form $\tilde{\Phi}_{\rm FD}=g_{\rm 1D}\tilde{\mu}^2$ where 
$g_{\rm 1D}$ 
is the density of quasi-one-dimensional states that become 
nested \cite{harrison6}. If we 
assume the limit $a_1\gg a_2$ and count only oscillatory terms, this 
free energy can be written as 
\begin{equation}\label{freeen}
    \tilde{\Phi}_{\rm FD}\approx-\frac{g_{\rm 1D}(a_1\hbar\omega_{\rm 
    c})^2}{2\pi^2}\cos{\bigg(\frac{4\pi pF}{B}\bigg)}. 
\end{equation}
The resulting frequency doubling term 
in the magnetisation, 
$\tilde{M}_{\rm FD}=-\partial\tilde{\Phi}_{\rm FD}/\partial B$, 
thus has the form 
\begin{equation}\label{MFD}
    \tilde{M}_{\rm FD}\approx -\frac{2a_1^2N\beta^\ast g_{\rm 1D}}
    {\pi g_{\rm 2D}}\sin\bigg(\frac{4\pi pF}{B}\bigg), 
\end{equation}
where $g_{\rm 2D}=N\beta^\ast/F$ 
is the total density of 2D states. Note that the sign of 
$\tilde{M}_{\rm FD}$ is 
negative and can have an amplitude as much as four times larger 
than the amplitude of the conventional dHvA contribution to the second 
harmonic. 
On inserting $g_{\rm 1D}/g_{\rm 2D}\sim$~1 into the expression 
\begin{equation}\label{ratio}
    \frac{\tilde{M}_{\rm FD}}{\tilde{M}_1}\approx
    -2a_1\frac{g_{\rm 1D}}{g_{\rm 2D}},
\end{equation}
for the harmonic ratio, the correct amplitude and sign of the second 
harmonic can be approximately reproduced in Fig. \ref{ampsCDW0} over a 
wide range of angles, with the exception of the interval 
15$^\circ<\theta<$~40$^\circ$. Only 
the FD model can account for the negative sign of the second harmonic 
over a wide range of orientations. While there exists some departure 
from the predictions of the FD model in the range 15$^\circ<\theta<$40$^\circ$, 
this is likely to be related to a similar anomaly in the amplitude of the 
fundamental over the same angular range.
\section{conclusion}
In summary,
we have shown that a single value of $\nu^\ast_0g$ can account for the 
field-orientation of the sign and phase of the dHvA oscillations 
deep within both the CDW$_0$ and CDW$_x$ 
phases above and below $B_{\rm k}$. The implications of this are 
twofold: first, the split waveform that occurs 
within the CDW$_0$ phase for field orientations $|\theta|<$~42$^\circ$
cannot be attributed to 
spin-splitting, and second, the role of {\it e-e} interactions appears not 
to change between the two phases. 

We have also shown that the field orientation dependence of the waveform 
within the CDW$_x$ phase is entirely consistent with the 
predictions for a canonical ensemble of electrons with a background reservoir 
of quasi-one-dimensional states. 
The behaviour of the waveform within the CDW$_0$ 
phase, however, is more unusual. The waveform of the 
oscillations appears to be significantly reduced in amplitude 
uniformly across the harmonics. Since this damping is not indexed to 
the harmonics, we can eliminate both impurity scattering and magnetic 
breakdown as dominant mechanisms for the reduction of the amplitude 
within the CDW$_0$ phase. Rather, it appears to be the case that the 
effective volume of the sample contributing to the dHvA signal is 
field dependent within the CDW$_0$ phase.

It is shown that the negative sign of the second harmonic that occurs 
within the CDW$_0$ phase over a large range of angles cannot be 
explained in terms of the dHvA effect. A negative sign is 
expected to follow naturally from the frequency doubling effect, 
however. The presence of frequency doubled oscillations in
$\alpha$-(BEDT-TTF)$_2$KHg(SCN)$_4$ is consistent with the existence 
of a commensurate 
CDW ground state \cite{harrison2}.

\section{acknowledgements}
The work is supported by the Department of Energy, the National 
Science Foundation (NSF) and the State of Florida. One of us (JSB), 
acknowledges the provision of an NSF grant (DMR-99-71474) and
(LB) would
like to acknowledge the provision of a FSU visiting scientist 
scholarship.

\begin{figure}
\caption{Examples of the oscillatory magnetic torque at several 
different angles measured in $\alpha$-(BEDT-TTF)$_2$KHg(SCN)$_4$ at  
450~$\pm$~20~mK throughout. The traces have been offset with respect 
to each other for clarity.
}
\label{wiggles}
\end{figure}

\begin{figure}
\caption{Fourier transform of the data at 
$\theta\sim$~8.8$^\circ$ in Fig. \ref{wiggles} 
over a restricted range of field (18.2~$<B<$~23~T). 
}
\label{transforms}
\end{figure}

\begin{figure}
\caption{(a) Field orientation-dependence of the Fourier amplitudes 
at different magnetic fields in $\alpha$-(BEDT-TTF)$_2$KHg(SCN)$_4$. 
Bezier fits between the points are shown for clarity. (b) The field 
orientation-dependence of the quantum oscillations at $B\sim$~26.5~T 
together with the functional form of $S_{1,\theta}\sin\theta$ best 
able to 
reproduce the correct positions of the nodes drawn as a solid line. 
(c) The 
field orientation-dependence of the quantum oscillations at 
$B\sim$~16.5~T, with the function form for $S_{1,\theta}\sin\theta$ 
shown with $\mu^\ast_0 g\sim$~3.67 (solid line) and 4.7 (dotted line)
respectively.
}
\label{phase}
\end{figure}

\begin{figure}
\caption{(a) An example of the oscillations in the magnetic torque 
measured in $\alpha$-(BEDT-TTF)$_2$KHg(SCN)$_4$ at low temperatures 
($T\sim$~27~mK) having subtracted the induced currents as described in 
Refernce [5]. (b) The calculated waveform of the oscillations, using 
the canonical ensemble as described in the text.
}
\label{waveform}
\end{figure}

\begin{figure}
\caption{(a) Field orientation-dependence of the amplitude of the 
fundamental $p=$~1 and second harmonic $p=$~2 in 
$\alpha$-(BEDT-TTF)$_2$KHg(SCN)$_4$ at 26.5~T, together with those 
calculated using to the canonical ensemble. (b) The same data but 
with the fundamental and second harmonic calculated using the grand 
canonical ensemble ({\it i.e.} the LK model).
}
\label{ampsCDWx}
\end{figure}

\begin{figure}
\caption{(a) Field orientation-dependence of the amplitude of the 
fundamental $p=$~1 and second harmonic $p=$~2 in 
$\alpha$-(BEDT-TTF)$_2$KHg(SCN)$_4$ at 16.5~T, together with those 
calculated using the canonical ensemble with $\Upsilon=$~79~T,
which equates to an effective scattering rate of 
2.9$\times$10$^{12}$~s$^{-1}$.
(b) The same data but with the fundamental and second harmonic 
calculated using the same scattering rate as within the CDW$_x$ phase, 
but with $\Upsilon^\prime=$~60~T.
}
\label{ampsCDW0}
\end{figure}


\begin{references}
\bibitem{note1} The quantity $\gamma$ represents the fraction of 
states in the material that compose the 2D $\alpha$ pocket. The other 
states are quasi-one-dimensional, functioning primarily as a charge 
reservoir.
\end{references}
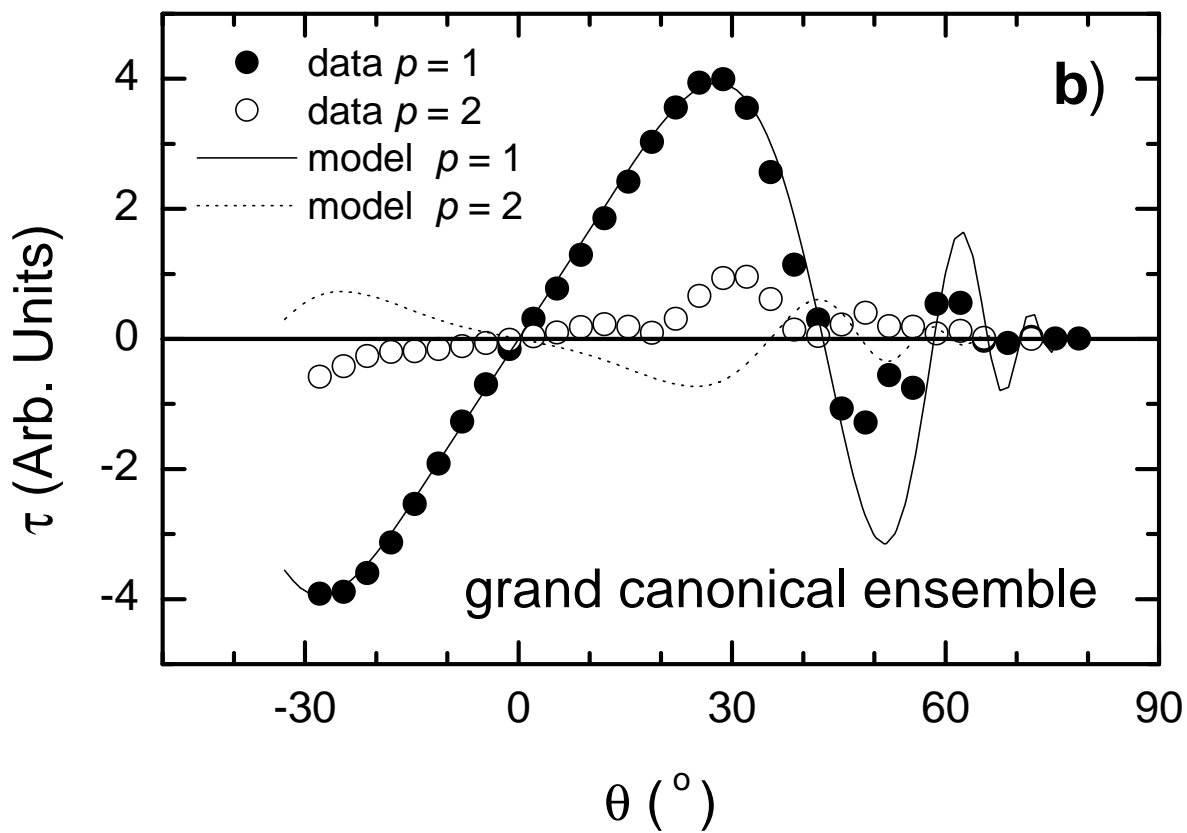
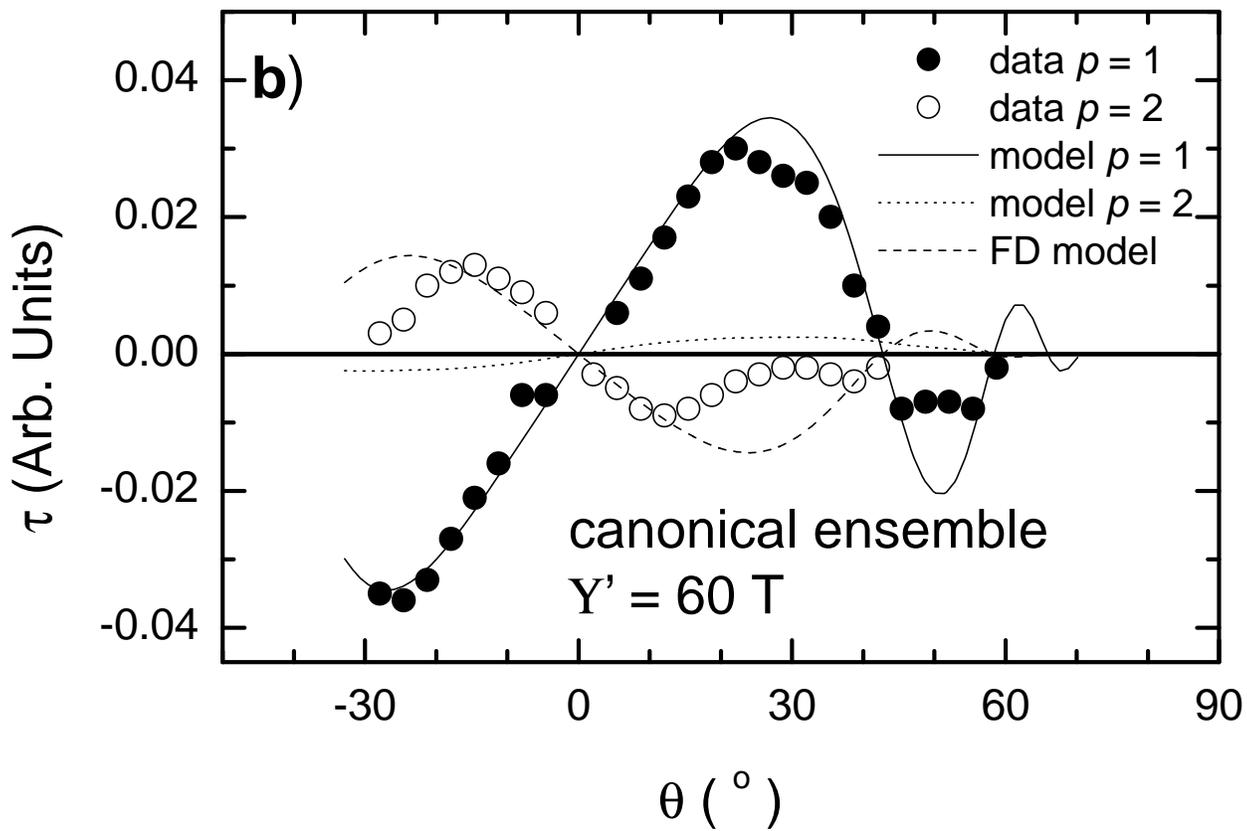
\end{document}